\documentclass[12pt]{article}
\let\section=\subsection     \let\subsection=\subsubsection                %%
\usepackage{graphicx}

\makeatletter
\def\lesssim{\mathrel{\mathpalette\vereq<}}
\def\vereq#1#2{\lower3pt\vbox{\baselineskip1.5pt \lineskip1.5pt
\ialign{$\m@th#1\hfill##\hfil$\crcr#2\crcr\sim\crcr}}}

\makeatother

\begin{document}
\begin{center}
   {\large \bf Recent Developments in Electromagnetic Excitation}\\[2mm]
   {\large \bf With Fast Heavy Ions}\\[5mm]
   G.~Baur$^a$\ \footnote{E-mail: g.baur@fz-juelich.de},
   K.~Hencken$^b$\ \footnote{E-mail: k.hencken@unibas.ch},
   D.~Trautmann$^b$\ \footnote{E-mail: dirk.trautmann@unibas.ch}
   and S.Typel$^c$\ \footnote{E-mail: s.typel@gsi.de}
 \\[5mm]
{\small \it $^a$Forschungszentrum J\"ulich, D-52425 J\"ulich, Germany\\
            $^b$Universit\"at Basel, CH-4056 Basel, Switzerland\\            
            $^c$GSI, D-64291 Darmstadt, Germany\\[8mm]}
\end{center}

\begin{abstract}\noindent
Coulomb dissociation is an especially simple and
important reaction mechanism. Since the
 perturbation due to the electric field 
of the (target) nucleus is exactly known, firm conclusions can be drawn
from such measurements. Electromagnetic matrixelements 
and astrophysical $S$-factors for radiative capture 
processes can be extracted from experiments. 
The dissociation 
of neutron halo nuclei
is studied in a zero range model using analytical methods.
% We
%discuss various approaches to the study of higher order 
%electromagnetic effects
%and how these effects depend on the basic parameters of the experiment.
%We also review ways how to treat nuclear interactions, 
%show their characteristics
%and how to avoid them (as far as possible). 
%We  review the experimental results
%from a theoretical point of view. 
Of special interest 
for nuclear structure physics is 
the appearence of low lying electric dipole strength in 
neutron rich nuclei. We use effective range methods to study it.
\end{abstract}
%\eject
%KH: not needed!!! \tableofcontents

\section{Introduction}

Electromagnetic excitation with heavy nuclei is a well established and
powerful 
tool in nuclear physics. With increasing beam energy higher lying
states like the giant dipole resonance can be excited and the nucleus
is readily dissociated. The theoretical description of these 
processes is given in  \cite{AlderW75,
BertulaniB88}. In the past years the field 
expanded a lot essentially  due to novel experiments done at 
intermediate energy radioactive beam facilities.
A recent review of the theoretical developments , along with 
a discussion of the experimental results can be found in \cite{BaurHT03}.
Applications to astrophysically relevant radiative capture reactions 
are also given there.

We would like to pick out a few issues in these conference 
proceedings. Part of the workshop was devoted to electron 
scattering, so it seems appropriate to highlight the similarities and 
differences of electromagnetic excitation with the weakly interacting
electron probe as compared to the excitation 
due to the strong field of a high Z nucleus.
This is done in Sect. \ref{elvshi}. In Sect. \ref{minmod}
 we study a simple model
(it is difficult enough, it is at least a three body problem) of 
the breakup of a halo nucleus bound by a zero range force in the 
pure Coulomb field of a nucleus. 
In Sect. \ref{effe} the effects of the 
finite range of the nuclear forces are studied.
 This is an appropriate 
approach since the spatial extension of halo nuclei
is larger than this range, thus providing a convenient expansion 
parameter. We use  effective range
methods.
An outlook and conclusions are presented in Sect.
\ref{sec:conclusions}.

\section{Electromagnetic Excitation with Electrons and Heavy
Ions, Similarities and Differences}
\label{elvshi} 
It is very important to note that 
with increasing beam energy higher lying states can be excited with
the Coulomb excitation mechanism. This can lead to 
Coulomb dissociation, in addition to Coulomb
excitation of particle-bound states. This was reviewed
some time ago in  \cite{BertulaniB88}.
It has become more and more clear, that 
such investigations are also well suited for secondary (radioactive) beams. 
An (unstable) fast projectile nucleus can interact with a 
high $Z$ target nucleus. In this way the interaction of an
unstable particle with a (quasireal or equivalent)
photon can be studied. A similar method is used in particle physics, 
where it is known as the Primakoff effect \cite{DreitleinP62,PomeranchuckS61}.

Let us give a very short primer of electromagnetic excitation with heavy 
ions pointing out differences and similarities to the 
excitation with electrons.
Since the electric field of a nucleus with high charge number $Z$
is much stronger than, e.g., the one of an electron, the nucleus 
can be a very suitable electromagnetic probe for certain cases. One can 
study, e.g., higher order phenomena,
which are inaccessible  with conventional electromagnetic probes like the
electron. The excitation of the double phonon giant 
dipole resonance 
at the GSI \cite{Emling94,AumannBE98} is an example. 

There are a few dimensionless parameters which characterize the  
electromagnetic excitation: we define the {\bf adiabaticity parameter}
as the ratio between the ``collision time'' 
and the ``excitation time'' 
\begin{equation}
\xi= \frac{\tau_{coll}}{\tau_{exc}}.
\end{equation}

We can estimate the collision time to be 
$\tau_{coll}=\frac{r_{min}}{\gamma v}$, 
where $r_{min}$ is the minimum impact parameter and 
 $\tau_{exc} =\frac{1}{\omega}$ , where $\omega$
is the nuclear excitation energy. From this we get
\begin{equation}
\xi=\frac{r_{min} \omega}{\gamma v}.
\label{eq:xidefined}
\end{equation}

For nonrelativistic collisions ($ v/c\ll 1$) we have $\gamma \approx 1$,
whereas in the relativistic case
the Lorentz parameter $\gamma$ can be much larger than one
and the collision time can 
become very small due to the Lorentz contraction.

For adiabaticity parameters $\xi \gg 1$ the system can 
follow the adiabatic ground state and no excitation occurs
\cite{Messiah85}.
This means also
that in nonrelativistic Coulomb excitation 
one can only excite nuclear states
for which the long-wavelength limit is valid:
due to the adiabaticity condition $\xi\lesssim 1$ we have
$\omega r_{min} \ll v$. This  leads to 
$k R \ll 1$, where $k=\frac{\omega}{c}$ and
$R$ denotes the size of the nucleus
% (see Fig.~\ref{fig_cxexplain}(a)), 
since $r_{min}>R$ and $v<c$.
On the other hand, for relativistic collisions the collision time
can be very small and one is able to excite states 
for which the long wavelength limit 
is no longer valid. 

A second parameter is the {\bf strength parameter},
which is defined as the strength of the interaction potential times its
duration(in units of $\hbar$):
\begin{eqnarray} 
\chi&=&\frac{V_{int}\tau_{coll}}{\hbar}.
\end{eqnarray} 
Here $V_{int}$ denotes a typical value of the interaction potential. 
For a multipole interaction of order $\lambda$, this value of the interaction
potential is of the order of 
$\gamma Z_1 e\left<f\right\|M(E\lambda)\left\|i\right>/r
_{min}^{\lambda+1}$
% (NR case)
, 
the strength parameter $\chi$ is therefore estimated to be  
\begin{equation}
\chi=\frac{Z_1 e<f||M(E\lambda)||i>}{\hbar v r_{min}^{\lambda}}.
\label{eq:chidefined}
\end{equation}
%(The relativistic case will be discussed below:
% in this case, the collision time 
%has a $\gamma^{-1}$-dependence. 
(The interaction $V_{int}$
is proportional to $\gamma$, the collision time 
$\tau_{coll}$ is inversely proportional to
$\gamma$  and thus $\chi$ becomes independent 
of $\gamma$.) 
The strength parameter for the monopole-monopole case is the 
Coulomb parameter, which is given by $\eta=\frac{Z_1 Z_2 e^2}{\hbar v}$.
The difference of heavy ions and electrons becomes obvious: the strength 
parameter is much larger for the heavy ions: this also means that higher order
effects are larger.

Due to the weak interaction the PWBA is 
the appropriate approximation for electron scattering and one has
a definite four-momentum transfer $q_{\mu}=
k_{\mu}-k'_{\mu}$.
%, that is, a definite energy transfer $\omega=q_0$ and
%a definite three-momentum transfer 
%$\vec{q}=(\vec{q_T},q_l)$. 
To a good approximation $q_l$ is given by the  minimum momentum transfer  
$q_l=q_{min}=\frac{\omega}{v}$
for small angle scattering and small energy loss. The invariant mass
of the photon is $Q^2=-q^2=q_T^2+(\frac{\omega}{\gamma v})^2$.
We always have $Q^2>0$: the exchanged photon is {\it ``virtual''} (spacelike).

In contrast to this the photons which are exchanged in Coulomb
excitation contain a sum over virtual photon momenta; they conspire 
in such a way that only an electromagnetic matrix-element survives,
which corresponds to the interaction with a real photon $Q^2=0$
(for details see 
%appendix Sec.~\ref{sec:appendix} 
\cite {BaurHT03}). The important thing is that the nuclei do 
not touch each other and that the strong interaction between them 
can be neglected.

\section{A Realistic Model for the Coulomb Dissociation of a 
Neutron Halo Nucleus}
\label{minmod}
Breakup processes in nucleus-nucleus collisions are complicated, in whatever
way they are treated. They constitute at least a three-body problem,
which is further complicated due to the long range Coulomb force. 
Exact treatments (like the Faddeev-approach) are therefore prohibitively 
cumbersome.
On the other hand, many approximate schemes 
have been developed in the field of direct nuclear reactions, and these 
approaches have been used with considerable success \cite{Austern70}. In this 
context we wish to investigate a realistic model for the Coulomb breakup 
of a neutron halo nucleus. 
 
We consider the breakup of a particle $a=(c+n)$ (deuteron, neutron-halo
nucleus) consisting of a loosely bound neutral particle $n$ and the core 
$c$ (with charge $Z_a=Z_c$) in the Coulomb field of a target nucleus with 
charge $Z$:
\begin{equation}
a+Z \rightarrow c+n+Z.
\label{eq:1}
\end{equation}
To simplify the kinematical relations we assume in this section
that the target is infinitely heavy.
We assume that the $a=(c+n)$ system is bound by a zero range force.
The potential $V_{cn}$ is adjusted to give one 
$s$-wave bound state with a binding energy $E_0$.
Neglecting the nuclear interaction of $c$ and $n$ with the target 
(``pure Coulomb'' case) the Hamiltonian of the system is given by
\begin{equation}
H=T_n + T_c + \frac{Z Z_c e^2}{r_c} + V_{cn}(r).
\end{equation}
The bound-state wave function of the system is given by 
%\begin{equation}
$
\phi_0 = \sqrt{\frac{\kappa}{2\pi}} \frac{\exp(-\kappa r)}{r},
$
%\label{eq:phigs}
%\end{equation}
where the quantity $\kappa$ is related to the binding energy $E_0$ of 
the system by
%\begin{equation}
$
E_{0} = \frac{\hbar^2 \kappa^2}{2 \mu},
$
%\end{equation}
and the reduced mass $\mu$ is given by
%\begin{equation}
$
\mu = \frac{m_n m_c}{m_n+m_c}.
$

The T-matrix in the Born approximation is found to be
\begin{equation}
T^{Born}=4 \pi \frac{Z Z_c e^2}{{\vec q}_{coul}\/^2} a_{fi}(\vec \Delta p),
\label{eq:tprior}
\end{equation}
where $\Delta p$ is related to the momentum transfer (or ``Coulomb push'')
to the target nucleus
\begin{equation}
\vec q_{coul}=\vec q_a-\vec q_c-\vec q_n
\label{eq:qcoul}
\end{equation}
by $ \vec \Delta p=\frac{m_n}{m_a} \vec q_{coul}$.
This ``Coulomb push'' has a perpendicular component $q_{coul\perp}$ 
and a component in the
beam direction $q_{coul\|}$. For high energies 
we have $q_{coul \|}=\frac{\omega}{v}$ (corresponding to the 
``minimum momentum transfer''). 
The amplitude $a_{fi}$ can be calculated analytically, see, e.g.,
Eqs.~33,34 of \cite{TypelB94a}.
It is found to be:
\begin{equation}
a_{fi}=\sqrt{8 \pi \kappa} (a_{FT} + a_S).
\end{equation}
The quantity $a_{FT}$ is essentially the Fourier transform
of the Yukawa wave function, given by 
\begin{equation}
a_{FT}= \frac{1}{(\vec q_{rel}-\vec \Delta p)^2+ \kappa^2},
\end{equation} 
where $q_{rel}$ denotes the relative momentum between c and n 
and $a_S$ takes the $s$-wave scattering part of the continuum wave into 
account:
\begin{equation}
a_S=\frac{i(\kappa+iq_{rel})}
{2|\Delta p|(\kappa^2+q_{rel}^2)} \ln{\frac{\kappa +i(q_{rel}+|\Delta p|}
{\kappa+i(q_{rel}-|\Delta p|)}}.
\end{equation}

 In the semiclassical
approach 
one calculates an impact parameter dependent breakup 
amplitude. It can be written as 
\begin{equation}
f_{breakup}=f_{coul} a_{fi},
\end{equation}
where $f_{coul}$ is the Rutherford amplitude.
The impact parameter $b$ is related to the ``Coulomb push'' by
the semiclassical relation
$b= \frac{2\eta_a}{q_{coul}}$.  
   The breakup amplitude in the sudden limit 
is given by (see eqs. 33 and 34 of \cite{TypelB94a})
$
a_{fi}=\left<q\right| \exp(i \vec \Delta p \vec r) \left| 0\right>.
$
One finds that the formula for the Born approximation is the
same as the one derived for the semiclassical sudden limit.
The ranges of validity of the two approaches, however,
do not overlap:
for the Born approximation we have $\eta_a\ll 1$ while the semiclassical
approximation requires $\eta_a \gg 1$.
In the sudden limit we have $\omega=0$ and there is only a transverse
momentum transfer $q_{coul\perp}$.
 
The Coulomb Distorted Wave Born Approximation is studied in \cite{BaurHT03}.
 At high beam energies 
it is found  that postacceleration effects disappear and that
(with minor well understood corrections) the quantal theory 
approaches the semiclassical straight line limit
for $\eta_a \gg 1$. For further
details we refer to this reference.

\section{Effective Range Theory of Halo Nucleus 
Photodissociation}
\label{effe}
Coulomb dissociation (or photodissociation, the time reversed process
of radiative capture) of halo nuclei shows some simple features.
Cross-sections as a function of energy tend to be universal,
when plotted in the appropriate reduced parameters.

Effective field theories are nowadays also used for 
the desription of halo nuclei, see \cite{BertulaniHK02}.
The relative momentum $k$
of the neutron and the core is indeed much smaller
than the inverse range of their interaction $1/R$ and
$k R$ is a suitable expansion parameter. (In our model of the pure-Coulomb
breakup of a bound state bound by a zero range force, 
see Sec.~\ref{minmod} 
above, we have $R=0$, i.e., $kR=0$ and we have the 
zero order contribution of the expansion). 
Effective range theory seems a natural starting point. This aspect was 
pursued in \cite{KalassaB96} 
and \cite{EsbensenBH97}.
In \cite{KalassaB96} radiative capture cross sections into $s$-,$p$- and 
$d$-bound
states are calculated in simple models, and the cross sections depend
only on a few low energy parameters. The neutron halo effect
on direct neutron capture and photodisintegration of $^{13}$C 
was studied in \cite{MengoniOI95} and \cite{OtsukaIFN94}.
In their figures it can very well be seen that the radial integrals are 
dominated by the outside region. While they find a sensitivity on neutron
optical model parameters for $s \rightarrow p$-capture, this sensitivity
is strongly reduced for the $p \rightarrow s$ 
and $p \rightarrow d$-capture cases.
In \cite{BertulaniHK02} it is remarked that the EFT approach ``is not 
unrelated to traditional single-particle models'' and that ``it remains to be 
seen whether these developments will prove to be a significant improvement 
over more traditional approaches.''  With a wealth of data on halo nuclei 
to be expected from the future rare ion beams we can be confident that 
these questions will be answered.

For the Coulomb breakup probability in an $ s \rightarrow p$ transition
we find:
\begin{eqnarray}
\frac{dP}{dq_{\rm rel}}&=&\frac{16y^2}{3\pi\kappa (1-\kappa r_0)}
% \nonumber & & 
% \times
 \left
( \cos \delta_1 \frac{x^2}{(1+x^2)^2} 
 + \sin\delta_1 \frac{1+3x^2}{2x(1+x^2)^3} \right)^2 \: .
\end{eqnarray}
where $r_0$ denotes the effective range parameter, the 
strength parameter y is defined in \cite{BaurHT03},
$\delta_1$ is the p-wave phase shift
 and $x=\frac{q_{rel}}{\kappa}$. In fig.1 we compare
the reduced transition probabilities corresponding to
this formula to a numerical calculation where  
a Woods-Saxon potential model is used. The agreement is very good.
For more strongly bound neutron-core systems and other
angular momentum states the effects due to the finite range
of the interaction
are expected to become more important. The present method
can also be extended to proton-core systems.
\begin{figure}
\unitlength1mm
\begin{picture}(0,65)
\put(5,-12){\makebox{\includegraphics[width=135mm]{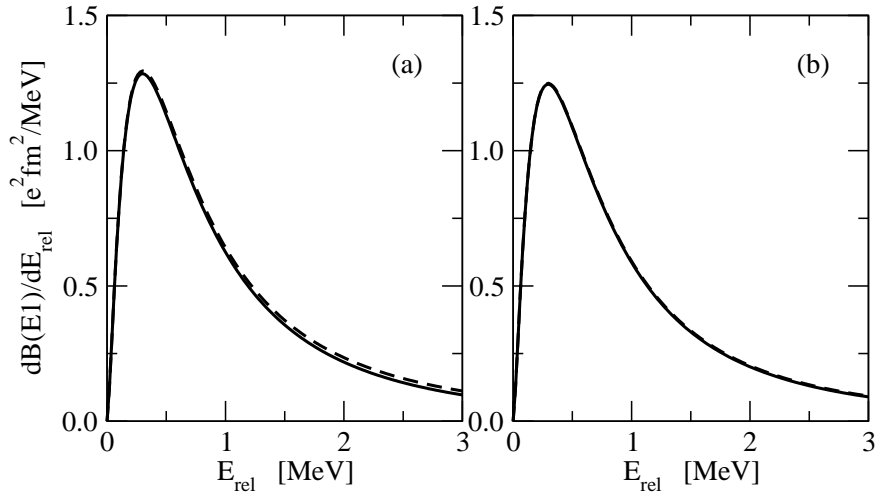}}}
\end{picture}
\caption{
Reduced transition probability as a function of the relative
energy for the breakup of ${}^{11}$Be into ${}^{10}$Be and a neutron
(a) with a plane wave in the final state and (b)
a scattering wave function
from a realistic potential. The solid line is the result of the
potential model calculation and the dashed line is the approximate
expression corrsponding to eq. 13.
}
\end{figure}

Light single particle 
halo nuclei, like $^{11}Be$ or $^{15,17,19}C$
have been studied with the Coulomb dissociation technique.
There are also studies of two-neutron halo nuclei,
notably $^{11}Li$. An effective range theory of these
two-neutron halo nuclei would be very interesting. An early work
related  to this field is  \cite{Migdal73}.
With the developments of the new radioactive beam
facilities applications to
medium and heavy one- and two-neutron halo nuclei
will become a promising field of study. 

%\input{intro}

%\input{theory}

%\input{higher}

%\input{analytic}

%\input{nuclear}

%\input{nstruc}

%\input{astro}

%\input{future}

%DO NOT CHANGE: $Id: conclusion.tex,v 1.8 2003/02/03 12:42:52 hencken Exp $
\section{Conclusions}
\label{sec:conclusions}

The intense source of quasi-real
(or equivalent) photons present in peripheral collisions
of medium and high energy nuclei (stable or radioactive)
has opened a wide horizon of related problems
and new experimental possibilities especially for the present and forthcoming 
radioactive beam facilities to investigate efficiently
photo-interactions with nuclei (single- and multiphoton excitations
and electromagnetic dissociation).
Let us mention the discovery of low lying 
E1 strength in neutron-rich nuclei and the determination
of astrophysical S-factors of radiative capture processes
like $^7$Be(p,$\gamma)^8$B.
The electromagnetic excitation of the giant dipole resonance at
the relativistic heavy ion colliders RHIC and LHC has also
become relevant. On the one hand,
these processes cause a decrease in luminosity,
on the other hand it they are
very useful as a luminosity monitor and a trigger on other
(even more interesting) processes in ultraperipheral collisions,
for an introduction and references see Ch. 6.2 of \cite{BaurHT03}.
 
\section{Acknowledgments}
We would like to thank 
C.~A.~Bertulani, H.~Rebel,
and R.~Shyam for their collaboration
on the present subject at 
various stages.

%\input{appendix}

%% Placing of figures:
%% If you have problems to place figures appropriately in the text with the
%% figure-environment, use this construction: 
%\begin{center}
%   \includegraphics[width=6cm,height=5cm,angle=-90]{figure_1.eps}\\
%   \parbox{14cm}
%        {\centerline{\footnotesize 
%        Fig.~1: Spectral function (left) of
%        the $\rho$-meson as a function \dots}}
%\end{center}
%\begin{thebibliography}{99}
%\itemsep=0cm
%\bibitem{GK}
%C.~Gale and J.~Kapusta, Phys.~Rev.~{\bf C35} (1987) 2107;
%\bibitem{KP}
%C.L.~Korpa and S.~Pratt, Phys.~Rev.~Lett.~{\bf 64} (1990) 1502.
%\end{thebibliography}
%\bibliography{physjabb,ppnp}

\begin{thebibliography}{10}

\bibitem{AlderW75}
K. Alder and A. Winther, {\em Electromagnetic excitation} (North-Holland,
  Amsterdam, 1975).

\bibitem{BertulaniB88}
C.~A. Bertulani and G. Baur, Phys. Rep. {\bf 163},  299  (1988).

\bibitem{BaurHT03}
G. Baur, K. Hencken, and D. Trautmann, Prog. Part. Nucl. Phys. {\bf 51},  487
  (2003).

\bibitem{DreitleinP62}
J. Dreitlein and H. Primakoff, Phys. Rev. {\bf 125},  1671  (1962).

\bibitem{PomeranchuckS61}
I.~Y. Pomeranchuck and I.~M. Shmushkevich, Nucl. Phys. {\bf 23},  452  (1961).

\bibitem{Emling94}
H. Emling, Prog. Part. Nucl. Phys. {\bf 33},  729  (1994).

\bibitem{AumannBE98}
T. Aumann, P.~F. Bortignon, and H. Emling, Annu. Rev. Nucl. Part. Sci. {\bf
  48},  351  (1998).

\bibitem{Messiah85}
A. Messiah, {\em Quantenmechanik} (Walter de Gruyter, Berlin, New York, 1985),
  Vol.~Band 2.

\bibitem{Austern70}
N. Austern, {\em Direct Reaction Theory} (Wiley, New York, 1970).

\bibitem{TypelB94a}
S. Typel and G. Baur, Nucl. Phys.~A {\bf 573},  486  (1994).

\bibitem{BertulaniHK02}
C.~A. Bertulani, H.-W. Hammer, and U. van Kolck, Nucl. Phys.~A {\bf 712},  37
  (2002).

\bibitem{KalassaB96}
D.~M. Kalassa and G. Baur, J. Phys.~G {\bf 22},  115  (1996).

\bibitem{EsbensenBH97}
H. Esbensen, G.~F. Bertsch, and K. Hencken, Phys. Rev.~C {\bf 56},  3054
  (1997).

\bibitem{MengoniOI95}
A. Mengoni {\it et~al.}, Phys. Rev.~C {\bf 52},  R2334  (1995).

\bibitem{OtsukaIFN94}
T. Otsuka {\it et~al.}, Phys. Rev.~C {\bf 49},  R2289  (1994).

\bibitem{Migdal73}
A.~B. Migdal, Sov. J. of Nucl. Phys. {\bf 16},  238  (1973).

\end{thebibliography}
%\bibliographystyle{prsty}

\end{document}